# MODELING STRONG SEISMIC GROUND MOTION:
# 3D LOADING PATH VS WAVEFIELD POLARIZATION

Maria Paola Santisi d'Avila[1], Luca Lenti[2] and Jean-François Semblat[2]

[1] Université Pierre et Marie Curie, IJLRDA, 75005 Paris, France. Email: paola.santisi_d_avila@upmc.fr
[2] Université Paris Est, IFSTTAR, 75015 Paris, France.

Corresponding author:
Maria Paola Santisi d'Avila
Université Pierre et Marie Curie
Institut Jean Le Rond D'Alembert

Address:
2, Rue du Pot de Fer
75005 Paris
France

Phone: +33 (0)1 44278708
Fax:   +33 (0)1 44275259
Email: paola.santisi_d_avila@upmc.fr
       mpaolasantisi@gmail.com

**SUMMARY**

Seismic waves due to strong earthquakes propagating in surficial soil layers may both reduce soil stiffness and increase the energy dissipation into the soil. In order to investigate seismic wave amplification in such cases, past studies have been devoted to one-directional shear wave propagation in a soil column (1D-propagation) considering one motion component only (1C-polarization). Three independent purely 1C computations may be performed ("1D-1C" approach) and directly superimposed in the case of weak motions (linear behavior). The present research aims at studying local site effects by considering seismic wave propagation in a 1D soil profile accounting for the influence of the 3D loading path and nonlinear hysteretic behavior of the soil. In the proposed "1D-3C" approach, the three components (3C-polarization) of the incident wave are simultaneously propagated into a horizontal multilayered soil. A 3D nonlinear constitutive relation for the soil is implemented in the framework of the Finite Element Method in the time domain. The complex rheology of soils is modeled by mean of a multi-surface cyclic plasticity model of the Masing-Prandtl-Ishlinskii-Iwan type. The great advantage of this choice is that the only data needed to describe the model is the modulus reduction curve. A parametric study is carried out to characterize the changes in the seismic motion of the surficial layers due to both incident wavefield properties and soil nonlinearities. The numerical simulations show a seismic response depending on several parameters such as polarization of seismic waves, material elastic and dynamic properties, as well as on the impedance contrast between layers and frequency content and oscillatory character of the input motion. The 3D loading path due to the 3C-polarization leads to multiaxial stress interaction that reduces soil strength and increases nonlinear effects. The nonlinear behavior of the soil may have beneficial or detrimental effects on the seismic response at the free surface, depending on the energy dissipation rate. Free surface time histories, stress-strain hysteresis loops and in-depth profiles of octahedral stress and strain are estimated for each soil column. The combination of three separate 1D-1C nonlinear analyses is compared to the proposed 1D-3C approach, evidencing the influence of the 3C-polarization and the 3D loading path on strong seismic motions.

# 1 INTRODUCTION

Numerous seismic records show that the local site condition is one of the dominant factors controlling the variation in ground motion and determination of the site-specific seismic hazard. Soils are complex materials and a linear approach is not reliable to model their seismic response to strong quakes. The evidence of nonlinear soil behavior comes from experimental cyclic tests on soil samples, for different strain amplitudes, where it is observed departure from the linear state as well as hysteresis when ground deformations up to around 0.01‰ are attained (Hardin & Drnevich 1972a; Hardin & Drnevich 1972b; Vucetic 1990). The nonlinearity is particularly manifested in shear modulus reduction and in the increase of damping for increasing strain levels. The effect on the transfer function of such nonlinear effects is a shift of the fundamental frequency toward lower frequencies, as well as an attenuation of the spectral amplitudes at high frequencies (Beresnev & Wen 1996). For places where recorded data are not available, but soil parameters are known, it is necessary to estimate theoretically the transfer function based on the parameters of the soil layers.

One-directional wave propagation analyses are an easy way to estimate the free surface ground motion, used as input signal in the design of structures. Schnabel et al. (1972) introduced the equivalent-linear analysis as a way to approximate the computation of nonlinear site response through an iterative procedure. In their method, the resulting shear modulus reduction and increasing damping are independent of the stress-strain path (Kramer 1996). Nevertheless, the popularity of the equivalent linear method is perhaps due to the small number of parameters needed, its ease of use and its rapidity compared to time domain wave propagation. The equivalent linear approach has been implemented into widely used codes, such as SHAKE (Schnabel et al. 1972) and EERA (Bardet et al. 2000) to investigate one-component ground response of horizontally layered sites. This method is assumed to be reasonable for strain levels between 0.01‰ and 1‰ (Ishihara 1996; Yoshida & Iai 1998). A complete nonlinear site response analysis with the incorporation of hysteresis appears to be fundamental to investigate local seismic effects for high strain levels. Furthermore, the three motion components are coupled due to the nonlinear behavior; they can not be computed separately.

A complete nonlinear analysis requires the propagation of a seismic wave in a nonlinear medium by integrating the wave equation in the time domain and using an appropriate constitutive model. Inputs to these analyses include acceleration time histories at bedrock and nonlinear material properties of the various soil strata underlying the site. The main difficulty in nonlinear analysis is to find a constitutive model that reproduces faithfully the nonlinear and hysteretic behavior of soil under cyclic loadings, with the minimum number of parameters. Realistic hysteretic behavior of soils is difficult to model because the yield surface may have a complex form. Some researchers adopt the theory of plasticity to describe the hysteresis of soil (Zienkiewicz et al. 1982; Chen & Baladi 1985; Chen & Mizuno 1990; Prevost & Popescu 1996; Ransamooj & Alwash 1997; Montans 2000); others propose simplified nonlinear models (Kausel & Assimaki 2002; Delépine et al. 2009) and other ones combine elasto-plastic constitutive equations with empirical rules (Ishihara & Towhata 1982; Finn 1982; Towhata & Ishihara 1985; Iai et al. 1990a; Iai et al. 1990b; Kimura et al. 1993). Classical empirical rules that describe the loading and unloading paths in the stress-strain space are the so-called Masing rules, presented in 1926, (Kramer 1996), that reproduce quite faithfully the hysteresis observed in the laboratory (Vucetic 1990). The main problem of these rules is that the computed stress may exceed the maximum strength of the material when an irregular load is applied (Pyke 1979; Li & Liao 1993). Several attempts have been done in order to overcome this difficulty (Pyke 1979; Vucetic 1990; Bonilla, 2000).

The nonlinear site response analysis allows following the time evolution of the stress and strain during seismic events and the resulting free surface ground motion. One-directional models for site response analysis are proposed by several authors (Joyner & Chen 1975; Joyner et al. 1981, Lee &

Finn 1978; Pyke 1979; Bonilla, 2000; Hartzell S. et al. 2004; Phillips & Hashash 2009). Furthermore, Li (1990) incorporates the three-dimensional cyclic plasticity soil model proposed by Wang et al. (1990) in a finite element procedure, in terms of effective stress, to simulate the one-directional wave propagation. However, this complex rheology needs an excessive number of parameters to characterize the soil model.

The nonlinear rheology used in the present research is a multi-surface cyclic plasticity mechanism that depends on few parameters that can be obtained from simple laboratory tests (Iwan 1967). Material properties include the dynamic shear modulus at low strain and the variation of shear modulus with shear strain. This rheology allows the dry soil to develop large strains in the range of stable nonlinearity. Because of its three-directional nature, the procedure can handle both shear wave and compression wave simultaneously and predict not only horizontal motion but vertical settlement too. Iwan's model is also called Masing-Prandtl-Ishlinskii-Iwan (MPII) model, according to Segalman & Starr (2008). Two years later Masing's postulate, in 1926, Prandtl proposed an elasto-plastic model with strain-hardening, re-examined by Ishlinskii in 1944, obtained by coupling a family of stops in parallel or of plays in series (Bertotti & Mayergoyz 2006). Segalman & Starr (2008) showed that for any material behavior which may described as a Masing model, there exists a unique parallel-series (strain based) Iwan system that provides forces as a function of the displacement history. The MPII formulation of soil hysteretic behavior can be used to examine case histories of well known stratigraphies as well as to investigate the role of critical parameters affecting the soil response.

In the present research, a finite element procedure to evaluate stratified level ground response to three-directional earthquakes is presented and the importance of the three-directional shaking problem is analyzed. The main feature of the procedure is that it solves the specific three-dimensional stress-strain problem with a one-directional approach.

The proposed "1D-3C" approach is implemented in a code called SWAP_3C (Seismic Wave Propagation - 3 Components). The implementation of the nonlinear cyclic constitutive model is presented in Sections 2 and 3. The code is then corroborated by comparison with the nonlinear finite difference code NERA (Bardet & Tobita 2001), for the unidirectional propagation of a one-component shear wave ("1D-1C"). The reliability of the proposed model is assessed and similar results are produced (Section 4). A parametric analysis is developed to understand the influence of a three-dimensional loading path and input polarization. The impact of a great vertical to horizontal peak acceleration ratio is investigated. Effects of soil and input properties in the dynamic response of soil columns are shown in Section 5. The conclusions are developed in Section 6.

## 2    IMPLEMENTATION OF THE NONLINEAR CONSTITUTIVE MODEL

The three components of the seismic motion are propagated into a multilayered column of nonlinear soil from the top of the underlying elastic bedrock, by using a finite element scheme. Along the horizontal direction, at a given depth, soil is assumed here to be a continuous, homogeneous and infinite medium. Soil stratification is discretized into a system of $N$ horizontal layers, parallel to the $xy$ plane, using quadratic line elements with three nodes (Fig. 1). Shear and pressure waves propagate vertically in $z$-direction. These hypotheses yield no strain variation in $x$- and $y$-direction. Transformations remain small during the process and the cross sections of three-dimensional soil elements remain planes.

Figure 1

According to a finite element modeling of a horizontally layered soil system, the strong form of equilibrium equation in dynamic analysis, including compatibility conditions, three-dimensional nonlinear constitutive relation and the imposed boundary conditions, is expressed in the matrix form as

$$\mathbf{M}\ddot{\mathbf{D}} + \mathbf{C}\dot{\mathbf{D}} + \mathbf{F}_{int} = \mathbf{F} \qquad (1)$$

where $\mathbf{M}$ is the mass matrix, $\dot{\mathbf{D}}$ and $\ddot{\mathbf{D}}$ are velocity and acceleration vectors, respectively, i.e. the first and second time derivatives of the displacement vector $\mathbf{D}$. $\mathbf{F}_{int}$ is the vector of nodal internal forces and $\mathbf{F}$ is the load vector. $\mathbf{C}$ is a damping matrix derived from the fixed absorbing boundary condition, as explained below. The Finite Element Method, as applied in the present research, is completely described in the works of Zienkiewicz (1971), Bathoz & Dhatt (1990), Reddy (1993) and Cook *et al*. (2002).

Discretizing the soil column into $n_e$ quadratic line elements and consequently into $n = 2n_e + 1$ nodes (Fig. 1), having three translational degrees of freedom each, yields a $3n$-dimensional displacement vector $\mathbf{D}$ composed by three blocks whose terms are the displacement of the $n$ nodes in $x$-, $y$- and $z$- direction, respectively. The assembled $(3n \times 3n)$-dimensional mass matrix $\mathbf{M}$ and the $3n$-dimensional vector of internal forces $\mathbf{F}_{int}$ result from the assemblage of $(9 \times 9)$-dimensional matrices like $\mathbf{M}^e$ and vectors $\mathbf{F}_{int}^e$, respectively, corresponding to the element $e$, which are expressed by

$$\mathbf{M}^e = \rho_e \int_0^{h_e} \mathbf{N}^T \mathbf{N} \, dz \qquad \mathbf{F}_{int}^e = \int_0^{h_e} \mathbf{B}^T \boldsymbol{\sigma} \, dz \qquad (2)$$

where $h_e$ is the finite element length and $\rho_e$ is the soil density assumed constant in the element. The terms of the 6-dimensional stress and strain vectors, defined as follows, are the independent stress and strain components, respectively:

$$\boldsymbol{\sigma} = \begin{bmatrix} \sigma_{xx} & \sigma_{yy} & \tau_{xy} & \tau_{yz} & \tau_{zx} & \sigma_{zz} \end{bmatrix}^T$$
$$\boldsymbol{\varepsilon} = \begin{bmatrix} \varepsilon_{xx} & \varepsilon_{yy} & \gamma_{xy} & \gamma_{yz} & \gamma_{zx} & \varepsilon_{zz} \end{bmatrix}^T \qquad (3)$$

where $\varepsilon_{xx} = 0$, $\varepsilon_{yy} = 0$ and $\varepsilon_{xy} = 0$, according to the hypothesis of infinite horizontal soil. In equation (2), $\mathbf{N}(z)$ is the $(3 \times 9)$-dimensional shape function matrix. Integrals in equation (2) are solved using the change of coordinates $z = (1+\zeta)h_e/2$ with $dz = h_e/2 \, d\zeta$, where $\zeta \in [-1,1]$ is the local coordinate in the element, and the Gaussian numerical integration. The shape function matrix is defined, in local coordinates, as

$$\mathbf{N}(\zeta) = \begin{bmatrix} N_1 & N_2 & N_3 & & & & & & \\ & & & N_1 & N_2 & N_3 & & & \\ & & & & & & N_1 & N_2 & N_3 \end{bmatrix} \qquad (4)$$

According to Cook *et al.* (2002), $N_1 = -\zeta(1-\zeta)/2$, $N_2 = (1-\zeta^2)$ and $N_3 = \zeta(1+\zeta)/2$ are the quadratic shape functions corresponding to the three-node line element used to discretize the soil column. The terms of the $(6\times 9)$-dimensional matrix $\mathbf{B}(z)$ are the spatial derivatives of the shape functions, according to compatibility conditions and to the hypothesis of no strain variation in the horizontal directions $x$ and $y$. If the strain vector is defined as $\boldsymbol{\varepsilon} = \partial \mathbf{u}$ (Cook *et al.* 2002), where the terms of $\mathbf{u}$ are the displacements in $x$-, $y$- and $z$-direction and $\partial$ is a matrix of differential operators defined in such a way that compatibility equations are verified, matrix $\mathbf{B} = \partial \mathbf{N}$ results like

$$\mathbf{B} = \begin{bmatrix} \mathbf{0}_3 & \mathbf{0}_3 & \mathbf{0}_3 & \mathbf{0}_3 & \mathbf{B}_z & \mathbf{0}_3 \\ \mathbf{0}_3 & \mathbf{0}_3 & \mathbf{0}_3 & \mathbf{B}_z & \mathbf{0}_3 & \mathbf{0}_3 \\ \mathbf{0}_3 & \mathbf{0}_3 & \mathbf{0}_3 & \mathbf{0}_3 & \mathbf{0}_3 & \mathbf{B}_z \end{bmatrix}^T \quad (5)$$

where $\mathbf{0}_3$ is a 3-dimensional null vector and $\mathbf{B}_z = [\partial N_1/\partial z \;\; \partial N_2/\partial z \;\; \partial N_3/\partial z]^T$ with $\partial N_i/\partial z = (\partial N_i/\partial \zeta)(\partial \zeta/\partial z)$ for $i = 1, 2, 3$ and $\partial \zeta/\partial z = 2/h_e$.

The assemblage of $(3n \times 3n)$-dimensional matrices and $3n$-dimensional vectors is independently done for each of the three $(n \times n)$-dimensional submatrices and $n$-dimensional subvectors, respectively, corresponding to $x$-, $y$- and $z$-direction of motion.

The system of horizontal soil layers is bounded at the top by the free surface and at the bottom by a semi-infinite elastic medium representing the seismic bedrock. The stresses normal to the free surface are assumed null and the following condition, implemented by Joyner & Chen (1975) in a finite difference formulation and used by Bardet & Tobita (2001) in NERA code, is applied at the soil-bedrock interface to take into account the finite rigidity of the bedrock:

$$-\mathbf{p}^T \boldsymbol{\sigma} = \mathbf{c}(\dot{\mathbf{u}} - 2\dot{\mathbf{u}}_b) \quad (6)$$

The stresses normal to the soil column base at the bedrock interface are $\mathbf{p}^T \boldsymbol{\sigma}$ and $\mathbf{c}$ is a $(3\times 3)$ diagonal matrix whose terms are $\rho_b v_{sb}$, $\rho_b v_{sb}$ and $\rho_b v_{pb}$. The parameters $\rho_b$, $v_{sb}$ and $v_{pb}$ are the bedrock density and shear and pressure wave velocities in the bedrock, respectively. The three terms of vector $\dot{\mathbf{u}}$ are the velocities in $x$-, $y$- and $z$-direction, respectively, at the interface soil-bedrock (node 1 in Fig. 1). The terms of the 3-dimensional vector $\dot{\mathbf{u}}_b$ are the input velocities, in the underlying elastic medium in directions $x$, $y$ and $z$, respectively. The boundary condition (6) allows energy to be radiated back into the underlying medium. According to equation (6), the damping matrix $\mathbf{C}^1$ and the load vector $\mathbf{F}^1$, for the first element $(e=1)$, are defined by

$$\mathbf{C}^1 = \int_0^{h_1} \mathbf{N}^T \mathbf{c} \mathbf{N} \, dz \qquad \mathbf{F}^1 = \int_0^{h_1} \mathbf{N}^T \mathbf{c}(2\dot{\mathbf{u}}_b) \, dz \qquad (7)$$

$\mathbf{C}^e$ and $\mathbf{F}^e$ are a null matrix and vector, respectively, for the other elements all over the soil profile. The minimum number of quadratic line elements per layer $n_e^j$ is defined considering that $p=10$ is the minimum number of nodes per wavelength to accurately represent the seismic signal (Kuhlemeyer & Lysmer 1973; Semblat & Brioist 2000) and it is evaluated as

$$\min n_e^j = \frac{H_j}{2} \frac{p f}{v_s} \qquad (8)$$

where $H_j$ is the thickness of layer $j$ (Fig. 1), $f$ is the frequency of the input signal and $v_s$ is the assumed minimum shear velocity in the medium, corresponding to a 70% reduction of the initial shear modulus. The seismic signal wavelength is equal to $v_s/f$.

The finite element model and the nonlinearity of soil require spatial and time discretization, respectively, to permit the problem solution. The rate type constitutive relation between stress and strain is linearized at each time step. Accordingly, equation (1) is expressed as

$$\mathbf{M} \Delta \ddot{\mathbf{D}}_k^i + \mathbf{C} \Delta \dot{\mathbf{D}}_k^i + \mathbf{K}_k^i \Delta \mathbf{D}_k^i = \Delta \mathbf{F}_k \qquad (9)$$

where the subscript $k$ indicates the time step $t_k$ and $i$ the iteration of the problem solving process, as explained below.

The stiffness matrix $\mathbf{K}_k^i$ is obtained by assembling $(9 \times 9)$-dimensional matrices as follows, with respect to element $e$:

$$k_k^{e,i} = \int_0^{h_e} \mathbf{B}^T \mathbf{E}_k^i \mathbf{B}\, dz \tag{10}$$

The tangent constitutive (6x6) matrix $\mathbf{E}_k^i$ is evaluated by the incremental constitutive relationship given by

$$\Delta \boldsymbol{\sigma}_k^i = \mathbf{E}_k^i \Delta \boldsymbol{\varepsilon}_k^i \tag{11}$$

According to Joyner (1975), the actual strain level and the strain and stress values at the previous time step allow to evaluate the tangent constitutive matrix $\mathbf{E}_k^i$ and the stress increment $\Delta \boldsymbol{\sigma}_k^i = \Delta \boldsymbol{\sigma}_k^i \left( \boldsymbol{\varepsilon}_k^i, \boldsymbol{\varepsilon}_{k-1}, \boldsymbol{\sigma}_{k-1} \right)$.

The step-by-step process is solved by the Newmark algorithm, expressed as follows:

$$\begin{cases} \Delta \dot{\mathbf{D}}_k^i = \dfrac{\gamma}{\beta \Delta t} \Delta \mathbf{D}_k^i - \dfrac{\gamma}{\beta} \dot{\mathbf{D}}_{k-1} + \left( 1 - \dfrac{\gamma}{2\beta} \right) \Delta t\, \ddot{\mathbf{D}}_{k-1} \\ \Delta \ddot{\mathbf{D}}_k^i = \dfrac{1}{\beta \Delta t^2} \Delta \mathbf{D}_k^i - \dfrac{1}{\beta \Delta t} \dot{\mathbf{D}}_{k-1} - \dfrac{1}{2\beta} \ddot{\mathbf{D}}_{k-1} \end{cases} \tag{12}$$

The Newmark procedure is a second-order approach for time integration in dynamic problems. The two parameters $\beta = 0.3025$ and $\gamma = 0.6$ guarantee a conditional numerical stability of the time integration scheme (Hughes 1987). Equations (9) and (12) yield

$$\overline{\mathbf{K}}_k^i \Delta \mathbf{D}_k^i = \Delta \mathbf{F}_k + \mathbf{A}_{k-1} \tag{13}$$

where the modified stiffness matrix is defined as

$$\overline{\mathbf{K}}_k^i = \frac{1}{\beta \Delta t^2} \mathbf{M} + \frac{\gamma}{\beta \Delta t} \mathbf{C} + \mathbf{K}_k^i \tag{14}$$

and $\mathbf{A}_{k-1}$ is a vector depending on the response in previous time step, given by

$$\mathbf{A}_{k-1} = \left[ \frac{1}{\beta \Delta t} \mathbf{M} + \frac{\gamma}{\beta} \mathbf{C} \right] \dot{\mathbf{D}}_{k-1} + \left[ \frac{1}{2\beta} \mathbf{M} + \left( \frac{\gamma}{2\beta} - 1 \right) \Delta t\, \mathbf{C} \right] \ddot{\mathbf{D}}_{k-1} \tag{15}$$

Equation (9) requires an iterative solving, at each time step $k$, to correct the tangent stiffness matrix $\mathbf{K}_k^i$. Starting from the stiffness matrix $\mathbf{K}_k^1 = \mathbf{K}_{k-1}$, evaluated at the previous time step, the value of matrix $\mathbf{K}_k^i$ is updated at each iteration $i$ (Crisfield 1991). After evaluating the displacement increment $\Delta \mathbf{D}_k^i$ by equation (13), using the tangent stiffness matrix corresponding to the previous time step, velocity and acceleration increments can be estimated through equation (12) and the total motion is obtained according to

$$\mathbf{D}_k^i = \mathbf{D}_{k-1} + \Delta \mathbf{D}_k^i \qquad \dot{\mathbf{D}}_k^i = \dot{\mathbf{D}}_{k-1} + \Delta \dot{\mathbf{D}}_k^i \qquad \ddot{\mathbf{D}}_k^i = \ddot{\mathbf{D}}_{k-1} + \Delta \ddot{\mathbf{D}}_k^i \tag{16}$$

where $\mathbf{D}_k^i$, $\dot{\mathbf{D}}_k^i$ and $\ddot{\mathbf{D}}_k^i$ are the vectors of total displacement, velocity and acceleration, respectively. The strain increment $\Delta \boldsymbol{\varepsilon}_k^i$ is then derived from the displacement increment $\Delta \mathbf{D}_k^i$. The stress increment $\Delta \boldsymbol{\sigma}_k^i$ and tangent constitutive matrix $\mathbf{E}_k^i$ are obtained through the constitutive relationship (11), according to the MPII approach. Gravity load is imposed as static initial condition in terms of strain and stress in nodes. The modified stiffness matrix $\bar{\mathbf{K}}_k^i$ is calculated and the process restarts. The correction process continues until the difference between two successive approximations is reduced to a fixed tolerance, according to

$$\left| \mathbf{D}_k^i - \mathbf{D}_k^{i-1} \right| < \alpha \left| \mathbf{D}_k^i \right| \tag{17}$$

where $\alpha = 10^{-3}$ (Mestat 1993; Mestat 1998). Afterwards, the next time step is analyzed.

The one-dimensional three-component propagation model ("1D-3C" approach) proposed in this Section is implemented in a code called SWAP_3C (Seismic Wave Propagation - 3 Components).

## 3  FEATURES OF THE CONSTITUTIVE MODEL

Modeling the propagation of a three-component earthquake in stratified soils requires a three-dimensional constitutive model for soil. The so-called Masing-Prandtl-Ishlinskii-Iwan (MPII) constitutive model, as suggested by Iwan (1967) and applied by Joyner (1975) and Joyner & Chen (1975) in a finite difference formulation, is used in the present work to properly model the nonlinear soil behavior in a finite element scheme. The MPII model is used to represent the behavior of materials satisfying Masing criterion (Kramer 1996) and not depending on the number of loading cycles. The stress level depends on the strain increment and strain history but not on the strain rate.

Therefore, this rheological model has no viscous damping. The energy dissipation process is purely hysteretic and does not depend on the frequency. Iwan (1967) proposed an extension of the standard incremental theory of plasticity (Fung 1965), by introducing a family of yield surfaces, modifying the 1D approach with a single yield surface in stress space. He models nonlinear stress-strain curves using a series of mechanical elements, having different stiffness and increasing sliding resistance. The MPII model takes into account the nonlinear hysteretic behavior of soils in a three-dimensional stress state, using an elasto-plastic approach with hardening, based on the definition of a series of nested yield surfaces, according to von Mises' criterion. Shear modulus and damping ratio are strain-dependent. The MPII hysteretic model for dry soils, used in the present research, is applied for strains in the range of stable nonlinearity.

The main feature of the MPII rheological model is that the only necessary input data, to identify soil properties in the applied constitutive model, is the shear modulus decay curve $G(\gamma)$ versus shear strain $\gamma$. The initial elastic shear modulus $G_0 = \rho v_s^2$, measured at the elastic behavior range limit $\gamma \cong 0.001‰$ (Fahey 1992), depends on the mass density $\rho$ and the shear wave velocity in the medium $v_s$. The P-wave modulus $M = \rho v_p^2$, depending on the pressure wave velocity in the medium $v_p$, characterizes the longitudinal behavior of soil. The $v_p/v_s$ ratio, evaluated by

$$(v_p/v_s)^2 = 2(1-\nu)/(1-2\nu) \tag{18}$$

is a function of the Poisson's ratio $\nu$. This is a parameter of the constitutive behavior for multiaxial load and of the interaction between components in the three-dimensional response.

In the present study the soil behavior is assumed adequately described by a hyperbolic stress-strain curve (Konder & Zelasko 1963; Hardin & Drnevich 1972b). This assumption yields a normalized shear modulus decay curve, used as input curve representing soil characteristics, expressed as

$$G/G_0 = 1/(1+|\gamma/\gamma_r|) \tag{19}$$

where $\gamma_r$ is a reference shear strain provided by test data corresponding to an actual tangent shear modulus equivalent to 50% of the initial shear modulus. The applied constitutive model (Iwan 1967; Joyner & Chen 1975; Joyner 1975) does not depend on the hyperbolic backbone curve. It could incorporate also shear modulus decay curves obtained from laboratory dynamic tests on soil samples.

The deviatoric constitutive matrix $\mathbf{E}_d$ for a three-dimensional soil element is deduced according to Joyner (1975). The total constitutive matrix $\mathbf{E}$ in equation (11), such that $\Delta\mathbf{\sigma} = \mathbf{E}\Delta\mathbf{\varepsilon}$, is evaluated in the proposed method starting from $\mathbf{E}_d$, according to

$$\mathbf{E} = \mathbf{S}\mathbf{T} + \mathbf{E}_d (\mathbf{H} - \mathbf{T}) \tag{20}$$

where

$$\begin{aligned}
\mathbf{S} &= \mathrm{diag}\{3K_B \quad 3K_B \quad 0 \quad 0 \quad 0 \quad 3K_B\} \\
\mathbf{H} &= \mathrm{diag}\{1 \quad 1 \quad 1/2 \quad 1/2 \quad 1/2 \quad 1\} \\
\mathbf{T} &= [\mathbf{t} \quad \mathbf{t} \quad \mathbf{0}_6 \quad \mathbf{0}_6 \quad \mathbf{0}_6 \quad \mathbf{t}]
\end{aligned} \tag{21}$$

$\mathbf{t} = [1/3 \quad 1/3 \quad 0 \quad 0 \quad 0 \quad 1/3]^T$ and $\mathbf{0}_6$ is a 6-dimensional null vector. Vectors and matrices in equation (21) are deduced according to the definition of stress and strain vectors in equation (3). The equation (20) is derived considering that the constitutive matrix $\mathbf{E}_d$, obtained according to the Iwan procedure, allows evaluating the vector of deviatoric stress increments $\Delta\mathbf{s}$ knowing the vector of deviatoric strain increments $\Delta\mathbf{e}$, according to

$$\Delta\mathbf{s} = \mathbf{E}_d \Delta\mathbf{e} \tag{22}$$

where the deviatoric strain vector is defined as

$$\begin{aligned}
\Delta\mathbf{e} &= \begin{bmatrix} \Delta e_{xx} & \Delta e_{yy} & \Delta e_{xy} & \Delta e_{yz} & \Delta e_{zx} & \Delta e_{zz} \end{bmatrix}^T = \mathbf{H}\Delta\mathbf{\varepsilon} - \Delta\mathbf{\varepsilon}_m = \\
&= \begin{bmatrix} (\Delta\varepsilon_{xx} - \Delta\varepsilon_m) & (\Delta e_{yy} - \Delta\varepsilon_m) & \Delta\gamma_{xy}/2 & \Delta\gamma_{yz}/2 & \Delta\gamma_{zx}/2 & (\Delta e_{zz} - \Delta\varepsilon_m) \end{bmatrix}^T
\end{aligned} \tag{23}$$

and it corresponds to the following deviatoric stress vector:

$$\Delta \mathbf{s} = \begin{bmatrix} \Delta s_{xx} & \Delta s_{yy} & \Delta s_{xy} & \Delta s_{yz} & \Delta s_{zx} & \Delta s_{zz} \end{bmatrix}^T = \Delta \boldsymbol{\sigma} - \Delta \boldsymbol{\sigma}_m =$$
$$= \begin{bmatrix} (\Delta \sigma_{xx} - \Delta \sigma_m) & (\Delta \sigma_{yy} - \Delta \sigma_m) & \Delta \tau_{xy} & \Delta \tau_{yz} & \Delta \tau_{zx} & (\Delta \sigma_{zz} - \Delta \sigma_m) \end{bmatrix}^T$$
(24)

The volumetric strain $\varepsilon_m = (\varepsilon_{xx} + \varepsilon_{yy} + \varepsilon_{zz})/3$ corresponds to the mean stress $\sigma_m = (\sigma_{xx} + \sigma_{yy} + \sigma_{zz})/3$. The relationship between $\varepsilon_m$ and $\sigma_m$ (Joyner 1975), supposed as elastic, depends on the Bulk modulus $K_B$, according to

$$\Delta \sigma_m = 3 K_B \Delta \varepsilon_m \tag{25}$$

The vectors of mean stress and strain are respectively defined by

$$\begin{aligned}
\Delta \boldsymbol{\sigma}_m &= \mathbf{S} \Delta \boldsymbol{\varepsilon}_m = \begin{bmatrix} \Delta \sigma_m & \Delta \sigma_m & 0 & 0 & 0 & \Delta \sigma_m \end{bmatrix}^T \\
\Delta \boldsymbol{\varepsilon}_m &= \mathbf{T} \Delta \boldsymbol{\varepsilon} = \begin{bmatrix} \Delta \varepsilon_m & \Delta \varepsilon_m & 0 & 0 & 0 & \Delta \varepsilon_m \end{bmatrix}^T
\end{aligned} \tag{26}$$

Equation (22) corresponds to $(\Delta \boldsymbol{\sigma} - \Delta \boldsymbol{\sigma}_m) = \mathbf{E}_d (\mathbf{H} \Delta \boldsymbol{\varepsilon} - \Delta \boldsymbol{\varepsilon}_m)$, according to (23) and (24). The equation (20) is consequently deduced according to (26) and (11).

The three-component ground motion is characterized by the modulus which is a unique scalar parameter. Similarly, octahedral shear stress (respectively strain) is chosen to combine the three-dimensional stress (respectively strain) components in a unique scalar parameter. It allows an adequate comparison of the simultaneous propagation of the three motion components (1D-3C) and the independent propagation of the three components (1D-1C) superposed a posteriori. The 1D-1C approach is a good approximation in the case of low strains within the linear range (superposition principle; Oppenheim *et al*. 1997). The effects of axial-shear stress interaction in multiaxial stress states have to be taken into account for higher strain rates, in the nonlinear range. Stress and strain rate in the one-dimensional (1D) soil profile due to the propagation of a three-component earthquake are therefore expressed in the following analysis in terms of octahedral shear stress and strain, respectively obtained by

$$\tau_{oct} = \frac{1}{3}\sqrt{\left(\sigma_{xx}-\sigma_{yy}\right)^2+\left(\sigma_{yy}-\sigma_{zz}\right)^2+\left(\sigma_{zz}-\sigma_{xx}\right)^2+6\left(\tau_{xy}^2+\tau_{yz}^2+\tau_{zx}^2\right)}$$

$$\gamma_{oct} = \frac{1}{3}\sqrt{2\left(\varepsilon_{zz}\right)^2+6\left(\varepsilon_{yz}^2+\varepsilon_{zx}^2\right)}$$

(27)

according to the hypothesis of infinite horizontal soil $\left(\varepsilon_{xx}=0, \varepsilon_{yy}=0, \varepsilon_{xy}=0\right)$.

# 4 ANALYSIS OF THE LOCAL 1D-1C SEISMIC RESPONSE

Four soil profiles are modeled in the present study consisting of three layers on seismic bedrock (Fig. 2). The shear wave velocity profile of these soil columns is deduced using the approach proposed by Cotton et al. (2006), based on the model of Boore & Joyner (1997). An average shear wave velocity in the upper 30m $V_{s30}$ of 350 - 300 - 250 m/s is assumed for columns A, B and C, respectively. Soil column D has not an increasing shear velocity with depth but the middle layer is the most rigid instead (Table 1).

Figure 2

Table 1

The reference shear strain $\gamma_r$ of the hyperbolic model (equation (19)) is assumed equal to 0.35 - 0.5 - 1‰ in the various cases. The Poisson's ratio is imposed equal to 0.3 - 0.4 - 0.45 to obtain different values of the wave velocity ratio $v_p/v_s$ in the medium (1.87 - 2.45 - 3.32, respectively).

The physical properties assumed for the bedrock are the density $\rho_b = 2100\,kg/m^3$, the shear velocity in the bedrock $v_{sb} = 1000\,m/s$ and the pressure wave velocity $v_{pb}$ is deduced by (18), by imposing a Poisson's ratio of 0.4.

The cyclic input signal used for the parametric analysis developed in the present research is the following Mavroeidis-Papageorgiou wavelet (Semblat & Pecker 2009) with a phase shift:

$$\ddot{u}(t) = \frac{\ddot{u}_{max}}{2}\left[1+\cos\left(\frac{2\pi f}{n_p}\left(t-\frac{t_f}{2}\right)\right)\right]\cos\left(2\pi f\left(t-\frac{t_f}{2}\right)\right)$$

(28)

where $\ddot{u}_{max}$, $f$ and $t_f = n_p/f$ are the signal amplitude, fundamental frequency and duration, respectively, and $n_p$ is the number of peaks that describes the oscillatory character of the motion.

Such simple wavelets classically allow an easier verification for various amplitudes and number of cycles. The interest of this specific wavelet (28) is justified in order to control both the predominant frequency and the number of cycles. The latter is very important due to the influence of the loading history. Various input signal parameters are chosen to assess their effects on the seismic ground motion. Input frequency $f$ is assumed equal to 2 - 3 - 5 - 7 Hz. The number of peaks $n_p$ in the time history is chosen equal to 5 - 10 - 20. Acceleration signals are halved to take into account the free surface effect and integrated, to obtain the corresponding input data in terms of vertically incident velocities, before being forced at the base of the horizontally multilayered soil profile. Various input polarizations are chosen assuming the acceleration component in $z$-direction $\ddot{u}_z$ equal to 0.1 - 0.7 - 0.8 - 0.9 times the acceleration component in $x$-direction $\ddot{u}_x$ and $\ddot{u}_y = \ddot{u}_x$ for all cases.

In the case of the one-component input, the nonlinear site response in time domain obtained by the proposed model is corroborated by comparison with output data acquired by the nonlinear code NERA (Bardet & Tobita 2001). NERA is a 1D-1C ground response analysis software where the one-component constitutive model suggested by Iwan (1967) is implemented in a finite difference formulation, using the boundary condition proposed by Joyner & Chen (1975). A Mavroeidis-Papageorgiou SH wavelet is considered with five peaks, $f = 3\,\text{Hz}$ and $\ddot{u}_{max} = 0.35\,\text{g}$, where $\text{g} = 9.81\,\text{m/s}^2$ is the gravitational acceleration. The proposed "1D-3C" approach (SWAP_3C code) is compared to NERA for a one-component input, propagated in the $z$-direction (Fig. 3). The reference shear strain $\gamma_r = 0.5‰$ is assumed uniform in the soil profile.

The one-directional dynamic response of the three multilayer soil columns A, B and C is analyzed in terms of maximum stress $\tau_{zx}$ and maximum strain $\varepsilon_{zx}$ profiles, hysteresis loop in the most deformed layer and free surface smoothed acceleration time histories. In the case of 1C propagation, the shear modulus decreases according to the shear modulus decay curve of the material. The stress-strain curve during a loading is referred to a backbone curve (Fig. 3), determined knowing the shear modulus decay curve. The obtained predictions are coherent with the evaluations obtained by

NERA, in terms of variation with depth of the maximum strain and stress, hysteresis loop and free surface acceleration (Fig. 3). Unwanted high frequencies in acceleration time-histories, derived from the numerical integration scheme, are suppressed by smoothing (Fig. 3c). Low-pass filtering could be more suitable for real signals. Free surface accelerations obtained by NERA are not altered.

## 5 PARAMETRIC ANALYSIS OF THE LOCAL 1D-3C SEISMIC RESPONSE

### 5.1 1D-3C vs 1D-1C approach

Modeling the propagation of a three-component earthquake in a soil column directly allows taking into account the interactions between shear and pressure components of seismic load in a one-directional seismic response analysis.

A cyclic signal is used to analyze nonlinear effects under a triaxial stress state. The dynamic response of a soil column to the propagation of a three-component signal is compared to the superposition of the three independently propagated components. Soil properties used in the present analysis are shown in Table 1. An input signal with $f = 3\,\text{Hz}$ and five peaks is imposed at the base of soil column B (Fig. 2). The reference shear strain $\gamma_r = 0.5‰$ is assumed uniform in the soil profile. The assumed PGA is equal to $0.35\,g$ for the two horizontal components and a ratio $\ddot{u}_z/\ddot{u}_x = 0.8$ is assumed for the vertical component. The dynamic response of soil column B is shown in Fig. 4.

Figure 4

Cyclic shear strains with amplitude greater than the elastic behavior range limit give open loops in the shear stress-shear strain plane, exhibiting strong hysteresis. The shear modulus decreases and the dissipation increases with increasing strain amplitude, due to nonlinear effects. The Fig. 4 shows the soil column cyclic response in terms of shear stress and strain in *x*-direction, when both it is affected by a triaxial input signal and the *x*-component of the input signal is independently propagated. From one to three components, for a given maximum strain amplitude, the shear modulus decreases and the dissipation increases. Under triaxial loading the material strength is lower than for simple shear loading referred to as the backbone curve. The compressive stiffness is also reduced for multiaxial loading.

The dynamic response to a 3C signal is represented in terms of modulus for acceleration, velocity and displacement time histories and in terms of octahedral parameters for stresses and strains. The modulus of acceleration at the outcropping bedrock, with a peak $\ddot{u}_{max} = 0.57\,\text{g}$, appears reduced at the free surface of the analyzed soil column for both 1D-1C and 1D-3C approaches (Fig. 4). Conversely, velocity modulus time histories are amplified. The interaction between multiaxial stresses in the 3C approach yields a reduction of the ground motion at the free surface. Maximum octahedral strain and stress profiles are obtained by (27) depending on profiles of maximum strains $\varepsilon_{zx}$, $\varepsilon_{yz}$, $\varepsilon_{zz}$ and maximum stresses $\tau_{zx}$, $\tau_{yz}$, $\sigma_{xx}$, $\sigma_{yy}$, $\sigma_{zz}$, respectively. Maximum strain and stress components are not simultaneous in the analyzed time history. The hysteresis loops in terms of octahedral strain and stress are obtained evaluating octahedral strain and stress time histories by equations (27), knowing $\varepsilon_{zx}$, $\varepsilon_{yz}$, $\varepsilon_{zz}$ and $\tau_{zx}$, $\tau_{yz}$, $\sigma_{xx}$, $\sigma_{yy}$, $\sigma_{zz}$, respectively, at each time step.

### 5.2 Influence of the soil properties

*Average shear wave velocity*

The 1D-3C dynamic response of columns A, B and C is compared in Fig. 5. The same input signal with five peaks, $f = 3\,\text{Hz}$, PGA equal to $0.35\,\text{g}$ for the two horizontal components and a ratio $\ddot{u}_{bz}/\ddot{u}_{bx} = 0.8$ is used. A reference shear strain $\gamma_r = 1‰$ and Poisson's ratio $\nu = 0.4$, assumed uniform in soil columns A, B and C, have been chosen to minimize nonlinear effects induced by lower $\gamma_r$ and Poisson's ratio. The most rigid profile A shows the largest strength and lowest strains. The opposite is obtained for the softest profile C. The free surface velocity is more amplified for the rigid column A and the higher rate of energy dissipation in a softer soil yields lower amplification in column C. The free surface acceleration is amplified in all analyzed soil columns, in this particular case, compared with the assumed acceleration peak $\ddot{u}_{max} = 0.57\,\text{g}$ at the outcropping bedrock (Fig. 5). Free surface velocity is similarly amplified.

<div style="text-align: right;">Figure 5</div>

*Reference shear strain*

Soil profile B is used to compare dynamic responses in the case of different reference shear strain $\gamma_r$ equal to 0.35 - 0.5 - 1‰, assumed uniform in all layers, with a Poisson's ratio $\nu = 0.4$. The input acceleration signal has five peaks, $f = 3\,\text{Hz}$, $\ddot{u}_{x\_max} = \ddot{u}_{y\_max} = 0.35\,\text{g}$ and $\ddot{u}_z/\ddot{u}_x = 0.8$. Nonlinear effects starting at a lower strain rate yield lower strength (Fig. 6) and lower velocity amplification at the free surface. An amplification of the acceleration signal is observed at the free surface for the case of $\gamma_r = 1‰$. The amplitude reduction is inversely related to the reference shear strain.

<div style="text-align: right;">Figure 6</div>

*Poisson's ratio*

An important variation in the dynamic response of soil profiles is observed for different values of the Poisson's ratio. The softest soil profile C is analyzed assuming a Poisson's ratio $\nu$ of 0.3 - 0.4 - 0.5 and a reference shear strain $\gamma_r = 0.35‰$, both supposed uniform in the soil profile. A lower value of $\nu$ yields a lower pressure to shear velocity ratio in the medium that causes greater signal amplification and multiaxial stress interaction, shown in hysteresis loops (Fig. 7).

<div style="text-align: right;">Figure 7</div>

Free surface acceleration appears amplified, compared to the signal at outcropping bedrock, for $\nu = 0.3$ and reduced for analyzed cases with $\nu$ greater than 0.4 (Fig. 7). Velocity time history is amplified in all investigated cases. The Fig. 7 shows a hysteresis loop in terms of shear stress and strain in $x$-direction with more obvious nonlinear effects and three-component interaction in the case of $\nu = 0.3$, rather than for higher values of the $v_p/v_s$ ratio.

Shear stress-strain cycles at 5m depth in column C, with $\gamma_r = 0.35‰$ and $\nu = 0.3$, are shown in Fig. 8 for the cases of horizontal PGA $\ddot{u}_{x\_max}$ equal to $0.35\,\text{g}$ (left) and $0.5\,\text{g}$ (right). The dynamic response in $x$-direction, influenced by the loading amplitude in directions $y$ and $z$ and by the lower pressure to shear velocity ratio in the soil, is assessed by the 1D-3C approach and, conversely, the 1D-1C scheme is not affected by the interaction of multiaxial stresses and strains. The loop shape changes in each cycle and this interaction effect increases with the PGA. The 1D-1C model does not permit to predict such change. The stress-strain cycles for each direction are altered as a consequence of the coupling between loading components, according to Montans' results (Montans 2000). This effect is more obvious for a low Poisson's ratio and increases with loading amplitude.

Figure 8

### 5.3 Seismic wave polarization and loading features

*Polarization*

The softest profile C is analyzed applying different input signals and comparing the dynamic response. Reference shear strain $\gamma_r = 0.35‰$ and Poisson's ratio $\nu = 0.4$ are assumed uniform in the soil profile. Soil column C is shaken by a three-component signal with equal component in $x$- and $y$-direction, with $\ddot{u}_{x\_max} = 0.35\,g$, $f = 3\,Hz$ and five peaks. The dynamic response to a signal with different ratio between $z$- and $x$-component $\left(\ddot{u}_z/\ddot{u}_x = 0.1 - 0.7 - 0.8 - 0.9\right)$ is shown in Fig. 9 to investigate the influence of a high bedrock pressure wave.

Velocity amplification increases with the $\ddot{u}_z/\ddot{u}_x$ ratio. The reduction of free surface acceleration, compared to the signal at outcropping bedrock, is greatly lowered by the increasing $\ddot{u}_z/\ddot{u}_x$ ratio. No significant differences are obtained in terms of maximum shear stress. The loop shape changes with increasing strain, as a consequence of the coupling between loading components, according to Montans (2000). This effect is less important in this case, with $\nu = 0.4$ (Fig. 9 for $\ddot{u}_z/\ddot{u}_x = 0.8$), than for $\nu = 0.3$ (Fig. 8). The hysteresis loop in terms of octahedral strain and stress confirms a three-component interaction effect with larger maximum octahedral strain and more obvious non linear behavior.

Figure 9

*Different fundamental frequency*

The two lowest natural frequencies of the soil profiles are $f = 2.96 - 7.01$ Hz for profile A, $f = 2.4 - 5.7$ Hz for profile B and $f = 2.0 - 4.8$ Hz for profile C. Based on this values four different input signals are propagated with fundamental frequency of 2 - 3 - 5 - 7Hz and five peaks. The horizontal PGA is $0.35\,\text{g}$ for both components and $\ddot{u}_z/\ddot{u}_x = 0.8$. Uniform reference shear strain $\gamma_r = 1‰$ and Poisson's ratio $\nu = 0.4$ are assumed in soil profiles A, B and C. The 2Hz signal yields highest strength and strains in soil column A, B (Fig. 10) and C.

The amplification of free surface velocity time history, compared with velocity at the outcropping bedrock, appears independent of frequency for all examined soil profiles (Fig. 10). Profiles A, B and C show different free surface acceleration amplitudes for each input signal at the soil-bedrock interface (Fig. 10). Acceleration signal with $f = 3\text{Hz}$ is the most amplified in soil column A, justified by the fact that the signal frequency is the closest to column fundamental frequency. The input signal with $f = 5\text{Hz}$, the nearest to second natural frequency of soil columns B and C, is the most amplified by these columns. The signals with $f = 2\text{Hz}$ is slightly amplified in column B and reduced in column C.

Figure 10

*Different number of cycles*

The number of cycles in the input signal affects the dynamic response of the various soil profiles. The energy dissipation rate is lower for five cycles and consequently a higher maximum strain is reached and larger amplification of velocity at the free surface is observed. The difference in acceleration deamplification is less obvious. More energy is dissipated with a greater number of cycles and lower strains are reached. The strength is not affected by the number of peaks of the input signal. The results obtained for profile B with homogeneous $\gamma_r = 0.5‰$ and $\nu = 0.4$, and for an input signal with $\ddot{u}_{x\_max} = \ddot{u}_{y\_max} = 0.35\,\text{g}$, $\ddot{u}_z/\ddot{u}_x = 0.8$ and $f = 3\text{Hz}$, are shown in Fig. 11.

Figure 11

## 5.4 Influence of the stratigraphic setting

*Reference shear strain variable with depth*

The reference shear strain has been assumed homogeneous in previous examples for simplification but, in general, soil columns do not have uniform shear strain corresponding to a 50% reduction of the shear modulus. The dynamic response of a soil profile like B in the case of homogeneous reference shear strain of 0.5‰ is compared with the case of different $\gamma_r$ in each layer, equal to 0.35‰, 0.5‰ and 1‰, for the surface, middle and bottom layer, respectively (Fig. 12). A Poisson's ratio $\nu = 0.4$ is assumed for the soil profile. A five peak signal with horizontal PGA of $0.35\,g$, a ratio $\ddot{u}_z/\ddot{u}_x = 0.8$ and $f = 3\,\text{Hz}$ is imposed at the soil-bedrock interface.

A lower reference shear strain $\gamma_r$ yields a significant nonlinear behavior for a lower strain rate. Higher strain and lower stress are observed in the surface layer ($\gamma_r = 0.35$‰), compared to the case of homogeneous $\gamma_r = 0.5$‰, and lower strains and higher stress in the bottom layer ($\gamma_r = 1$‰). At the free surface, the layer with lower $\gamma_r$ (variable reference shear strain profile) yields less amplification in the velocity time history and more reduction of the acceleration signal. The variation in the free surface signal is less obvious for the acceleration time history than for the velocity. If the profile B, with different reference shear strain for each layer, is compared with the case of homogeneous $\gamma_r$ equal to 0.35‰, hysteresis loops in the surface layer shows lower strains in the soil profile with homogeneous reference shear strain. Larger hysteresis loops in the surface layer are observed in the case of reference shear strain variable with depth, in spite of the larger reference shear strain in bottom layers.

Figure 12

*Intermediate stiff layer*

Soil strength increases with depth but, analyzing 3D stratigraphy of basins, soil columns with more rigid materials in surficial layers could be identified. The dynamic response at the surface soil layer of soil columns B and D are compared in Fig. 13. Strength increases with depth in soil profile B

(Fig. 12, variable $\gamma_r$ curve). The effect of a more rigid middle soil layer (profile D) is analyzed.

A reference shear strain $\gamma_r$ equal to 0.35‰, 0.5‰ and 1‰ is assumed for the first, second and third layer, respectively, with $\nu = 0.4$ for all layers. A five peak signal with horizontal PGA of $0.35\text{g}$, a ratio $\ddot{x}_z/\ddot{x}_x = 0.8$ and $f = 3\text{Hz}$ is imposed at soil column base. Lower strains and unmodified strength are observed in the surface layer of column D and higher strains and lower strength in the bottom layer. Profiles A, B and C have the same trend when compared with D. The free surface velocity and acceleration amplitude is slightly greater when the rigidity is not regularly increasing with depth.

# 6    CONCLUSIONS

A geomechanical model is proposed to analyze the one-dimensional propagation of seismic waves due to strong quakes and accounting for the three motion components (1D-3C approach). A finite element modeling of horizontally layered soil is proposed, by adopting a three-dimensional constitutive relation of the Masing-Prandtl-Ishlinskii-Iwan (MPII) type that needs few parameters to characterize the hysteretic behavior of soils.

The proposed method provides a promising solution for strong seismic ground motion evaluation and site effect analysis.

A parametric study is presented to evidence the effects of the input motion polarization and 3D loading path analyzed by the "1D-3C" approach. The combination of three separate "1D-1C" nonlinear analyses is compared to the proposed "1D-3C" approach.

Multiaxial stress states induce strength reduction of the material and larger damping effects. Soil properties such as the upper limit of linear behavior range and the Poisson's ratio have great impact in local seismic response, influencing the soil dissipative properties. Input motion properties such as polarization (vertical to horizontal component ratio), fundamental frequency and oscillatory character (number of peaks) affect energy dissipation rate and thus the amplification effect. In particular, a low wave velocity ratio in the soil and a high vertical to horizontal component ratio increase the three-dimensional mechanical interaction and progressively change the hysteresis loop size and shape at each cycle.

The proposed model is verified by comparison with a finite difference unidirectional one-component propagation model of the literature (NERA code). Validation of the "1D-3C" approach against recorded free surface time histories should now be carried out. Local site effects in the Tohoku area during the 2011 Tohoku earthquake are actually being investigated by the authors, using the one-dimensional three-component propagation model proposed in this paper.

The MPII hysteretic model, used in the present research for dry soils, is applied for strains in the range of stable nonlinearity. The extension of the proposed "1D-3C" approach to higher strain rates is planned as further investigation to be able to study the effects of soil nonlinearity in drained conditions.

The Finite Element Method efficiency when strong heterogeneities and complex geometries are modeled allows an extension of the present approach to 2D and 3D alluvial basins but the amount of data in both linear and nonlinear ranges would be huge.

## AKNOWLEDGMENTS

The first author is very grateful to Prof. Mauro Schulz for his constructive suggestions. This work was partly funded by the French National Research Agency (Quantitative Seismic Hazard Assessment research project) and by the Radioprotection and Nuclear Safety Institute.

**Figure 1**. Spatial discretization of a horizontally layered soil forced at its base by a three-component earthquake.

**Figure 2**. Dynamic response of four soil profiles representative of a multilayered horizontal soil shaken at their base by a halved signal, recorded at outcropping bedrock.

**Figure 3**. Comparison between results obtained by the proposed "1D-3C" approach (SWAP_3C code) and NERA for a one-component input in profiles A, B and C. a) Maximum strain (left) and stress (right) profiles; b) Shear stress-strain loop at 41 m depth; c) Free surface acceleration time history.

**Figure 4**. Comparison between dynamic responses of soil profile B to one- and three-component input signal. a) Free surface acceleration and velocity time history (modulus) compared with bedrock signal; b) Maximum octahedral strain and stress profiles; c) Octahedral stress-strain loop at 9m depth; d) Shear and normal stress-strain loops at 9m depth.

**Figure 5**. Influence of average shear velocity: dynamic response of soil profiles A, B and C to a three-component input signal. a) Maximum octahedral strain and stress profiles; b) Shear stress-strain loops at 9m depth; c) Free surface velocity (left) and acceleration (right) time history compared with the bedrock time history (modulus).

**Figure 6**. Effect of the reference shear strain: dynamic response to a 3C input signal of three soil profiles (B type) with homogeneous reference shear strain equal to 0.35 - 0.5 - 1‰. a) Free surface acceleration time history (modulus) compared with the bedrock acceleration; b) Maximum octahedral strain and stress profiles; c) Shear (left) and octahedral (right) stress-strain loops at 9m depth.

**Figure 7**. Effect of Poisson's ratio: dynamic response to a 3C input signal of three soil profiles (C type) with homogeneous Poisson's ratio equal to 0.3 - 0.4 - 0.45. a) Free surface acceleration time history (modulus) compared with bedrock acceleration; b) Maximum octahedral strain and stress profiles; c) Shear (left) and octahedral (right) stress-strain loops at 9m depth.

**Figure 8.** Shear stress-strain loop at 5m depth for soil profile C with $\nu = 0.3$, in the cases of horizontal PGA equal to $0.35\,g$ (left) and $0.5\,g$ (right).

**Figure 9.** Effect of seismic wave polarization: dynamic responses of soil profile C to a three-component input signal with $\ddot{u}_z/\ddot{u}_x$ ratio equal to 0.1 - 0.7 - 0.8 - 0.9%. a) Free surface velocity (left) and acceleration (right) time history compared with bedrock velocity and acceleration peaks, respectively; b) Shear (left) and octahedral (right) stress-strain loops at 5m depth.

**Figure 10**. Comparison between dynamic responses of three soil profiles to a 3C input signal with fundamental frequency equal to 2 - 3 - 5 - 7Hz. a) Maximum octahedral strain and stress profiles (B type); b) Shear stress-strain loop at 9m depth (B type); b) Free surface velocity (top) and acceleration (bottom) time history compared with bedrock velocity and acceleration peaks, respectively, for soil profiles A (left), B (middle) and C (right).

**Figure 11**. Comparison between dynamic responses of soil profile B to a three-component input signal with 5 - 10 - 20 cycles. a) Free surface velocity (left) and acceleration (right) signals compared with bedrock velocity and acceleration peaks, respectively; b) Shear (left) and octahedral (right) stress-strain loop at 9m depth.

**Figure12**. Comparison between the dynamic response to a three-component input signal of two soil profiles (B type) with homogeneous and variable reference shear strain. a) Maximum octahedral strain and stress profiles (left) and free surface acceleration time history compared with the bedrock acceleration (right); b) Maximum octahedral strain and stress profiles (left) and shear stress-strain loop at 9m depth (right).

**Figure 13**. Influence of a stiff middle layer: dynamic response to a three-component input signal of soil profiles B and D with variable reference shear strain. a) Maximum octahedral strain and stress profiles; b) Shear stress-strain loop at 9m depth.

**Table 1**. Stratigraphy of soil profiles and soil properties

|  | | A | B | C | D |
|---|---|---|---|---|---|
| z [m] | th [m] | $v_s$ [m/s] | $v_s$ [m/s] | $v_s$ [m/s] | $v_s$ [m/s] |
| 0 - 10 | 10 | 270 | 220 | 185 | 220 |
| 10 - 30 | 20 | 490 | 400 | 335 | 550 |
| 30 - 50 | 20 | 675 | 550 | 460 | 400 |

**Table 1**. Stratigraphy of soil profiles and soil properties

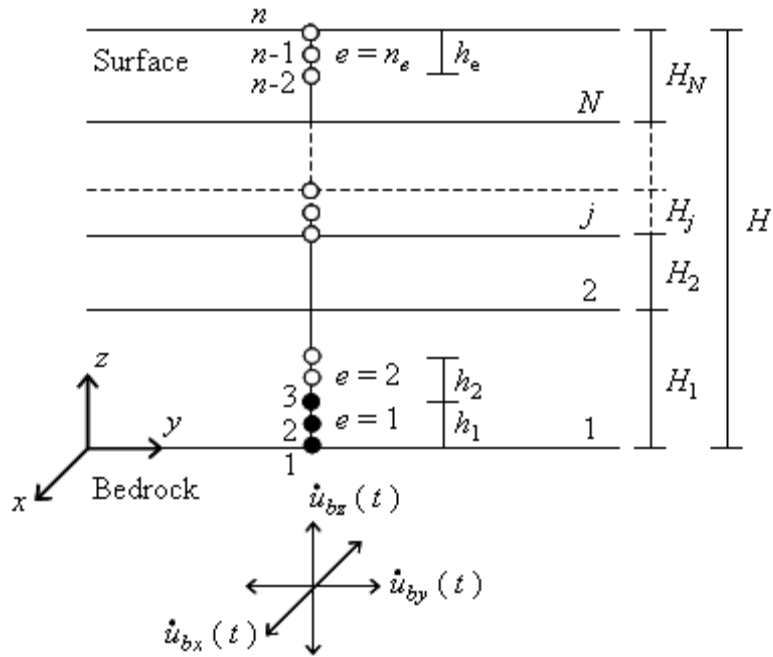

**Figure 1**. Spatial discretization of a horizontally layered soil forced at its base by a three-component earthquake.

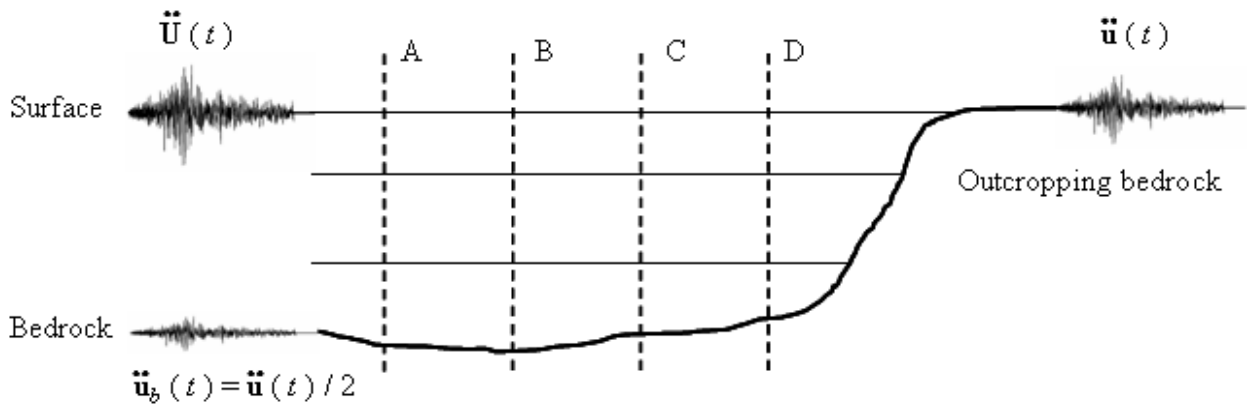

**Figure 2**. Dynamic response of four soil profiles representative of a multilayered horizontal soil shaken at their base by a halved signal, recorded at outcropping bedrock.

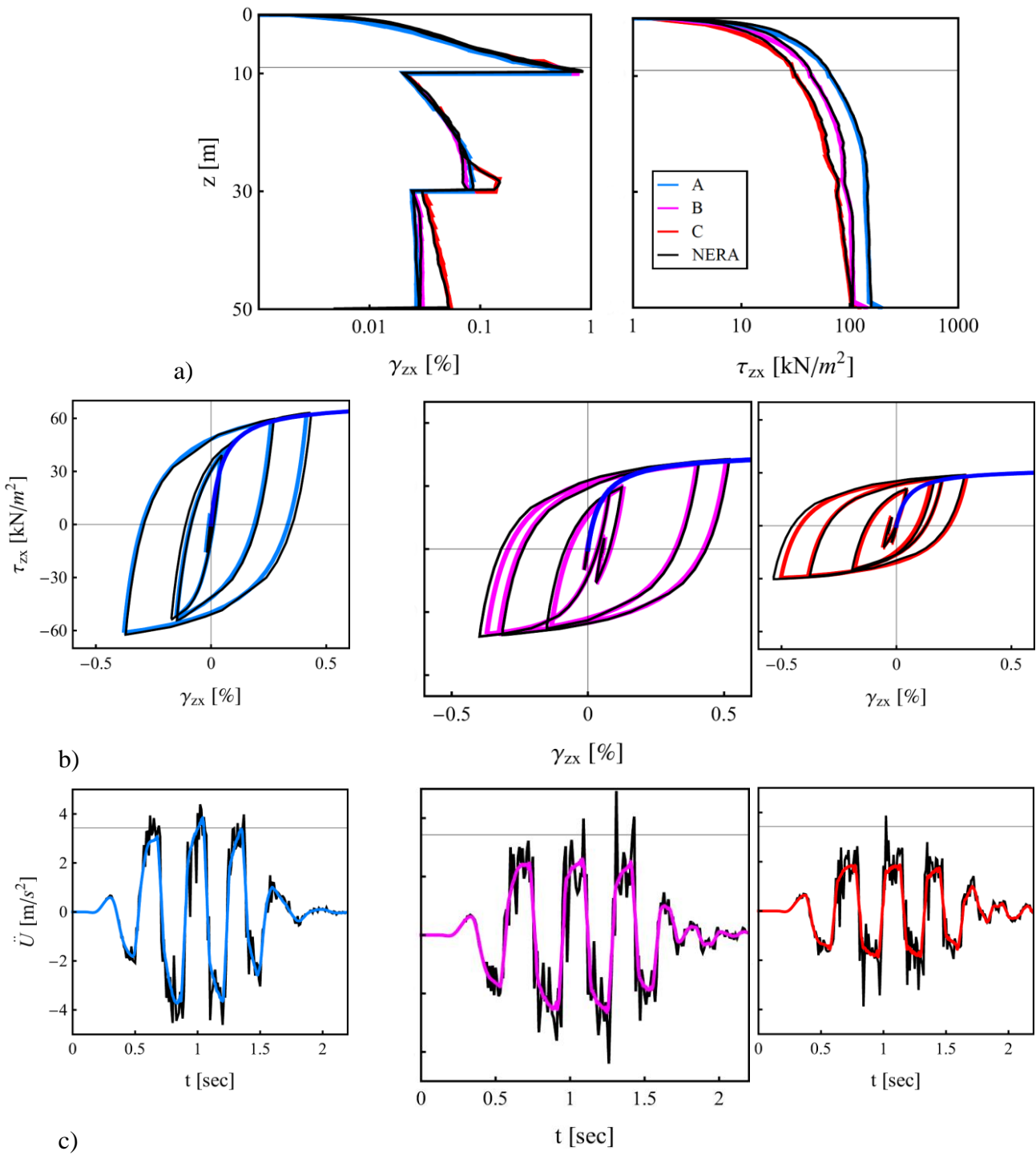

**Figure 3.** Comparison between results obtained by the proposed "1D-3C" approach (SWAP_3C code) and NERA for a one-component input in profiles A, B and C. a) Maximum strain (left) and stress (right) profiles; b) Shear stress-strain loop at 41 m depth; c) Free surface acceleration time history.

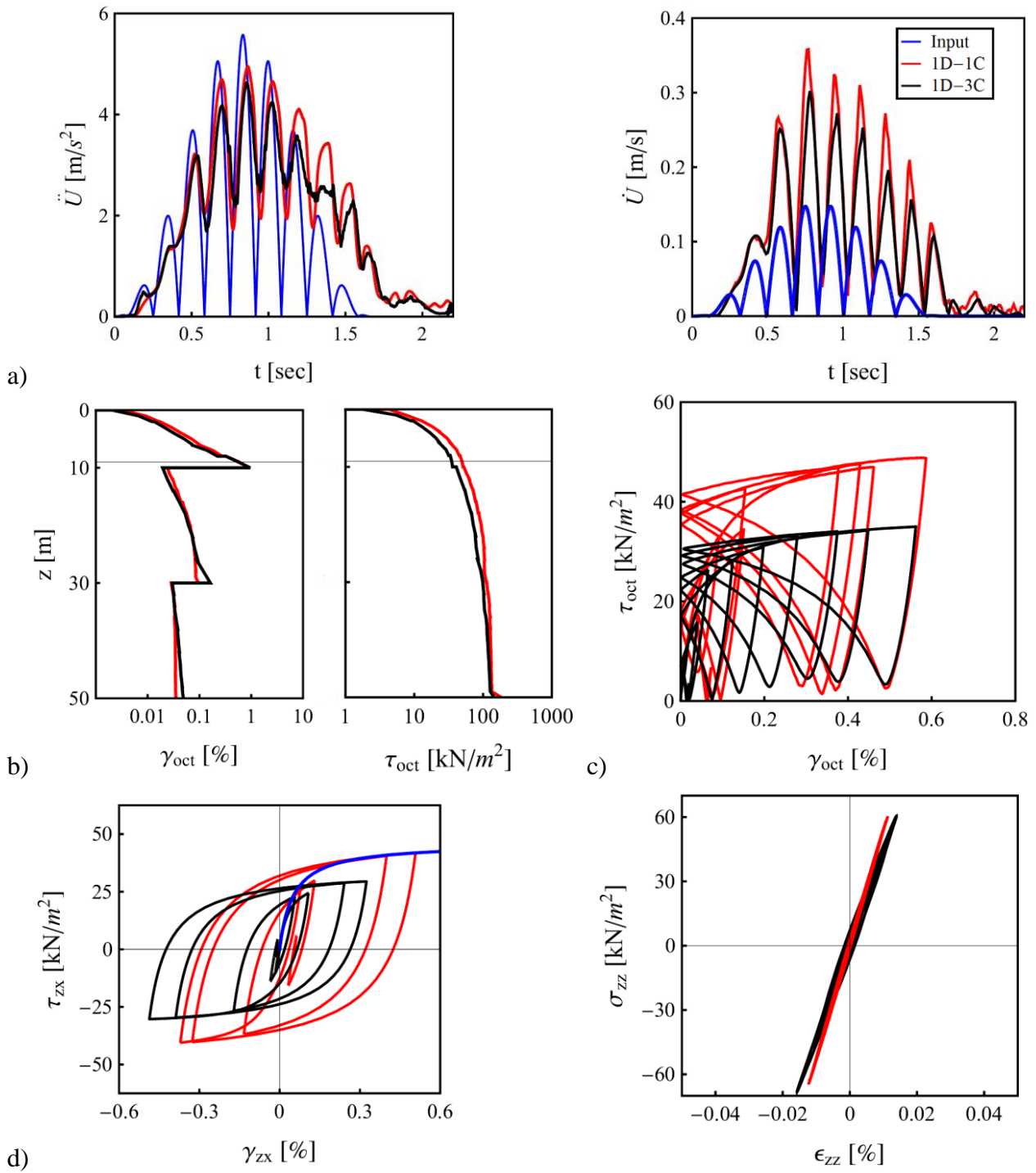

**Figure 4**. Comparison between dynamic responses of soil profile B to one- and three-component input signal. a) Free surface acceleration and velocity time history (modulus) compared with bedrock signal; b) Maximum octahedral strain and stress profiles; c) Octahedral stress-strain loop at 9m depth; d) Shear and normal stress-strain loops at 9m depth.

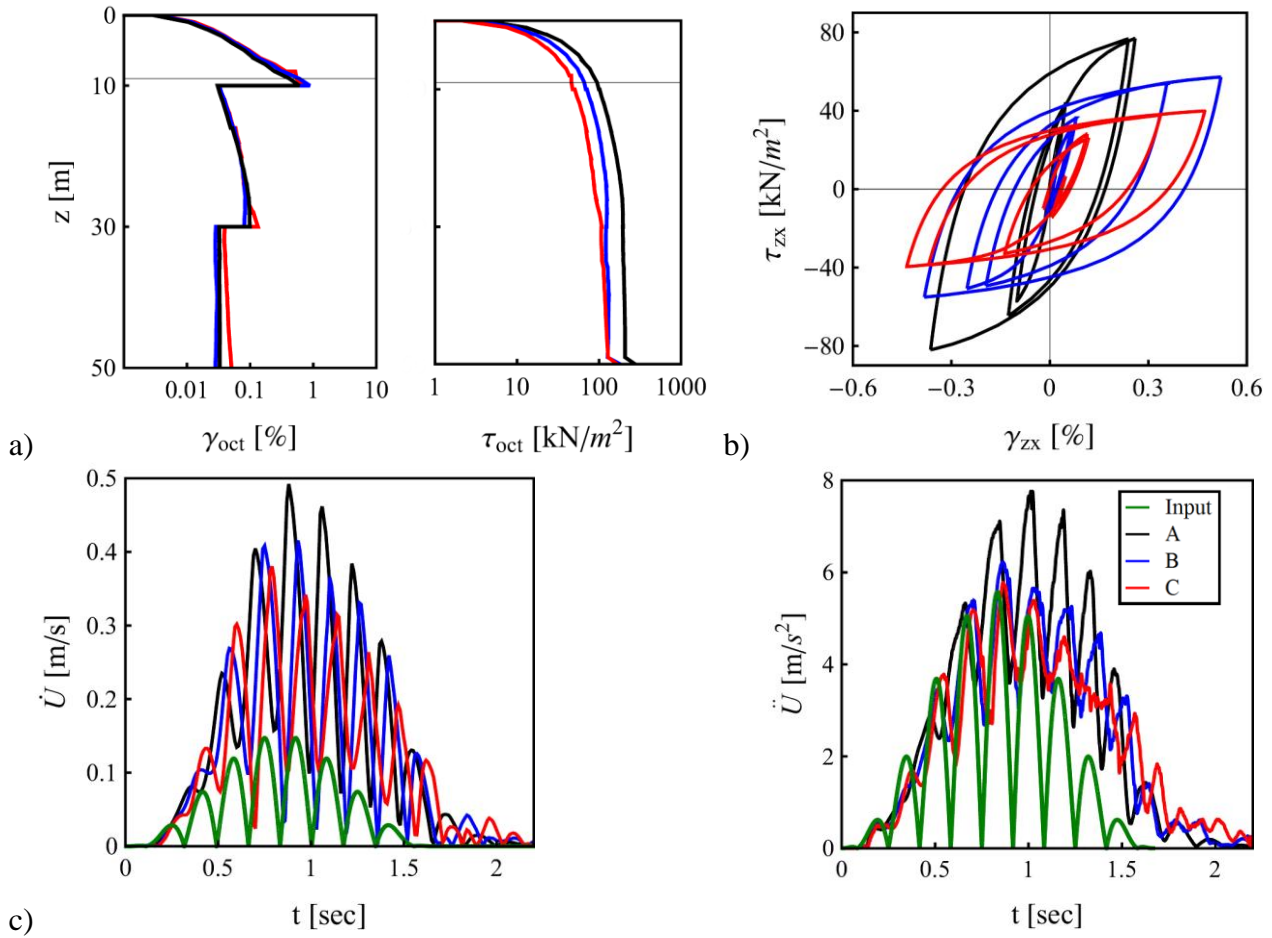

**Figure 5**. Influence of average shear velocity: dynamic response of soil profiles A, B and C to a three-component input signal. a) Maximum octahedral strain and stress profiles; b) Shear stress-strain loops at 9m depth; c) Free surface velocity (left) and acceleration (right) time history compared with the bedrock time history (modulus).

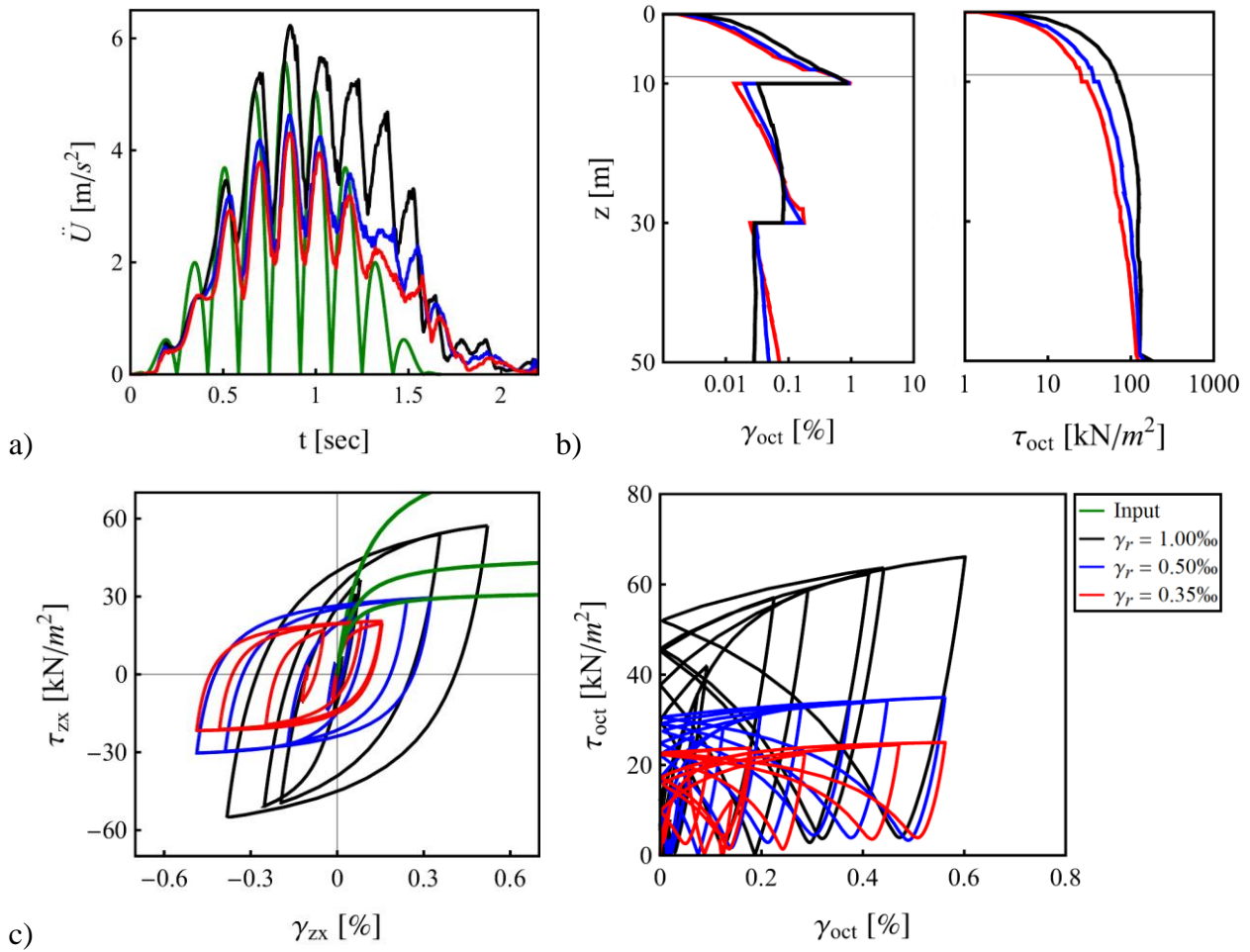

**Figure 6**. Effect of the reference shear strain: dynamic response to a 3C input signal of three soil profiles (B type) with homogeneous reference shear strain equal to 0.35 - 0.5 - 1‰. a) Free surface acceleration time history (modulus) compared with the bedrock acceleration; b) Maximum octahedral strain and stress profiles; c) Shear (left) and octahedral (right) stress-strain loops at 9m depth.

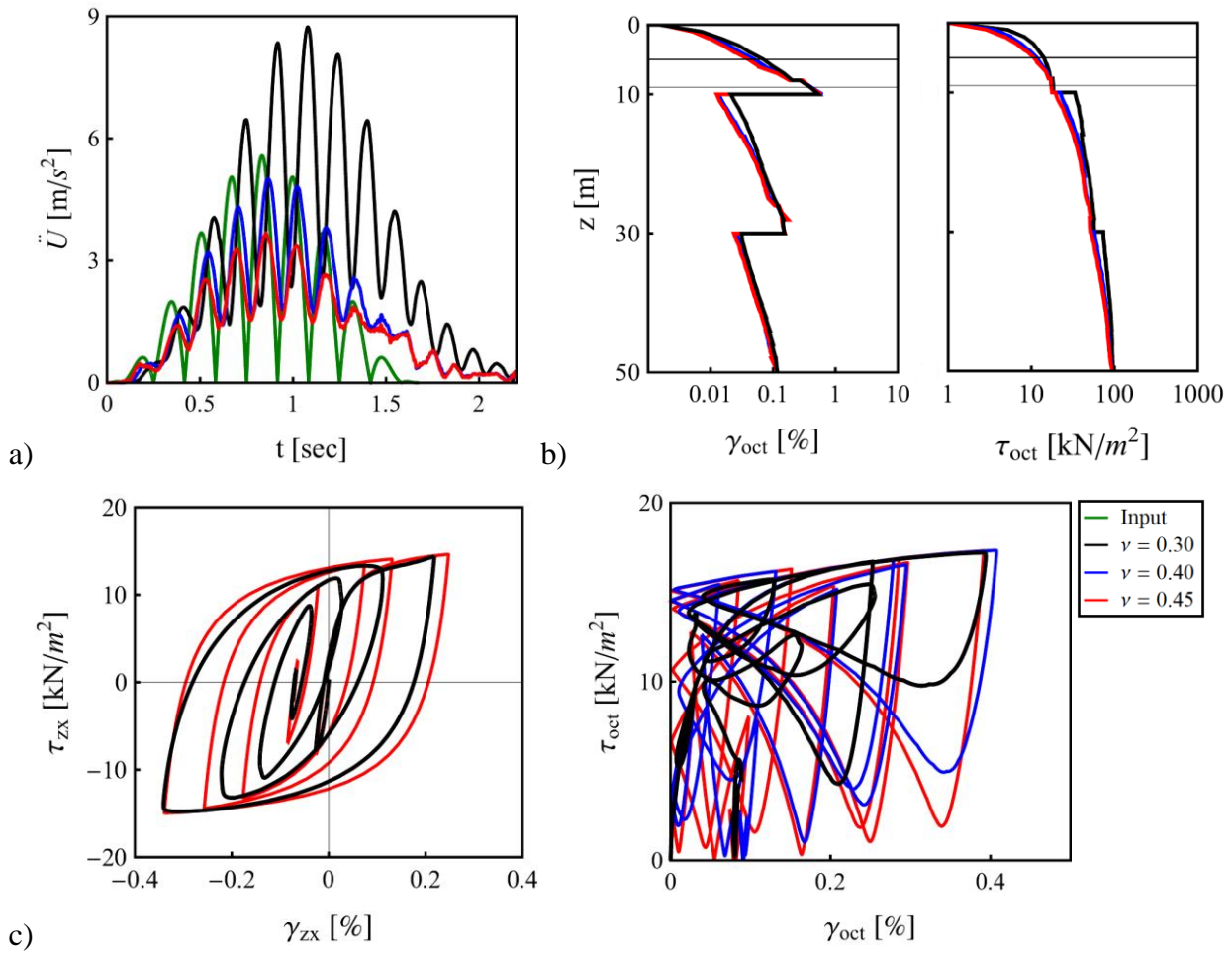

**Figure 7**. Effect of Poisson's ratio: dynamic response to a 3C input signal of three soil profiles (C type) with homogeneous Poisson's ratio equal to 0.3 - 0.4 - 0.45. a) Free surface acceleration time history (modulus) compared with bedrock acceleration; b) Maximum octahedral strain and stress profiles; c) Shear (left) and octahedral (right) stress-strain loops at 9m depth.

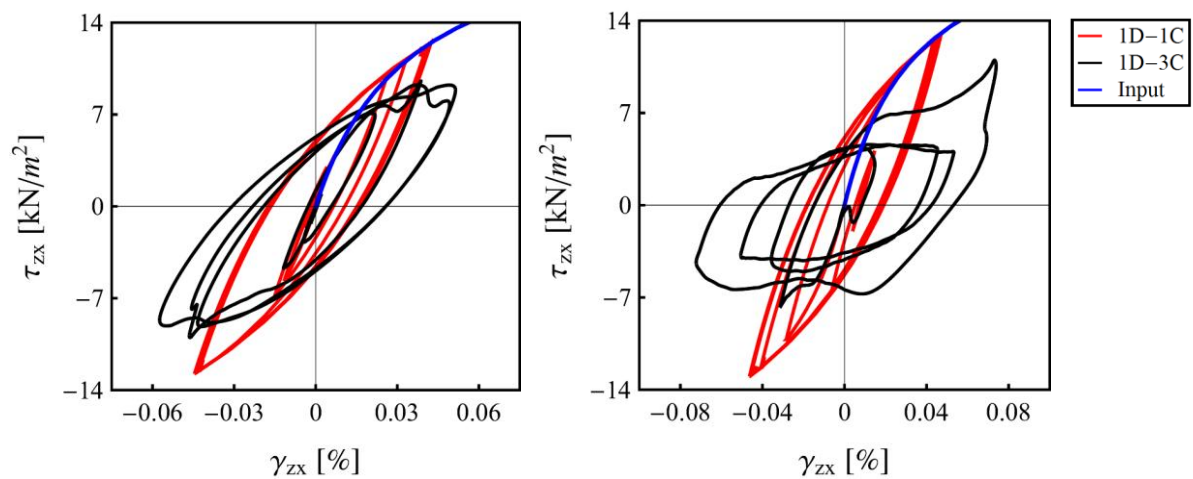

**Figure 8**. Shear stress-strain loop at 5m depth for soil profile C with $\nu = 0.3$, in the cases of horizontal PGA equal to $0.35\,g$ (left) and $0.5\,g$ (right).

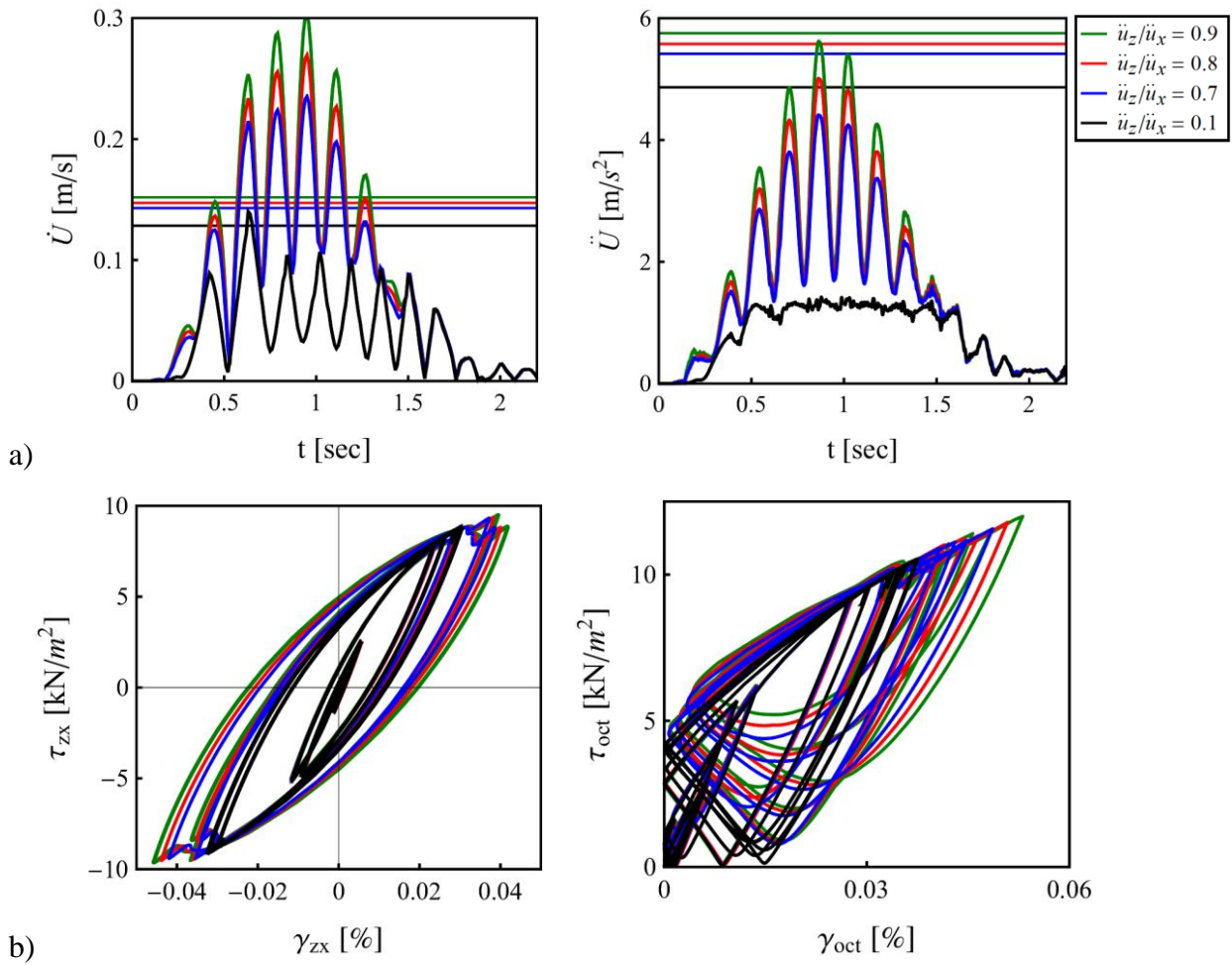

**Figure 9**. Effect of seismic wave polarization: dynamic responses of soil profile C to a three-component input signal with $\ddot{u}_z/\ddot{u}_x$ ratio equal to 0.1 - 0.7 - 0.8 - 0.9%. a) Free surface velocity (left) and acceleration (right) time history compared with bedrock velocity and acceleration peaks, respectively; b) Shear (left) and octahedral (right) stress-strain loops at 5m depth.

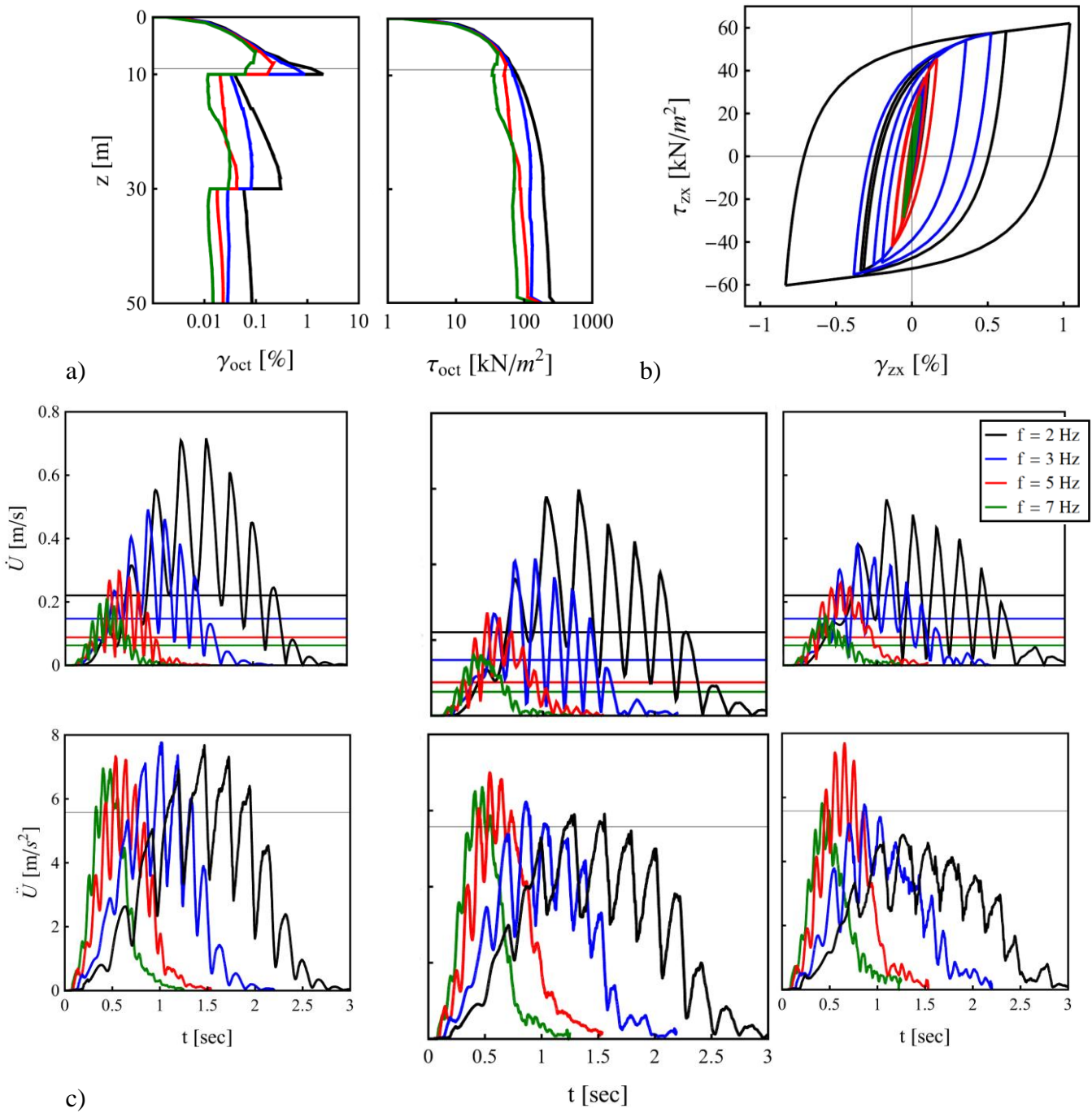

**Figure 10**. Comparison between dynamic responses of three soil profiles to a 3C input signal with fundamental frequency equal to 2 - 3 - 5 - 7Hz. a) Maximum octahedral strain and stress profiles (B type); b) Shear stress-strain loop at 9m depth (B type); b) Free surface velocity (top) and acceleration (bottom) time history compared with bedrock velocity and acceleration peaks, respectively, for soil profiles A (left), B (middle) and C (right).

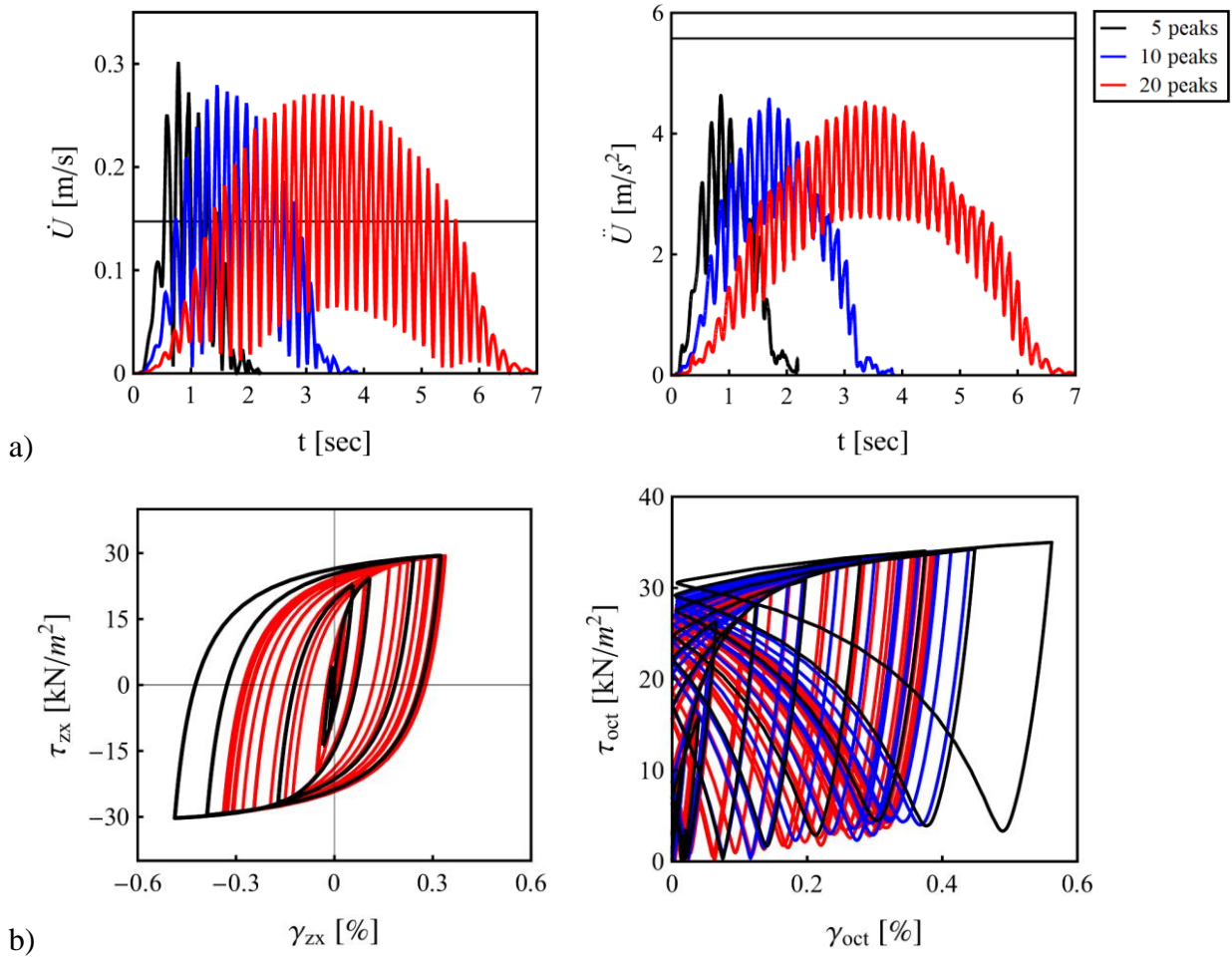

**Figure 11**. Comparison between dynamic responses of soil profile B to a three-component input signal with 5 - 10 - 20 cycles. a) Free surface velocity (left) and acceleration (right) signals compared with bedrock velocity and acceleration peaks, respectively; b) Shear (left) and octahedral (right) stress-strain loop at 9m depth.

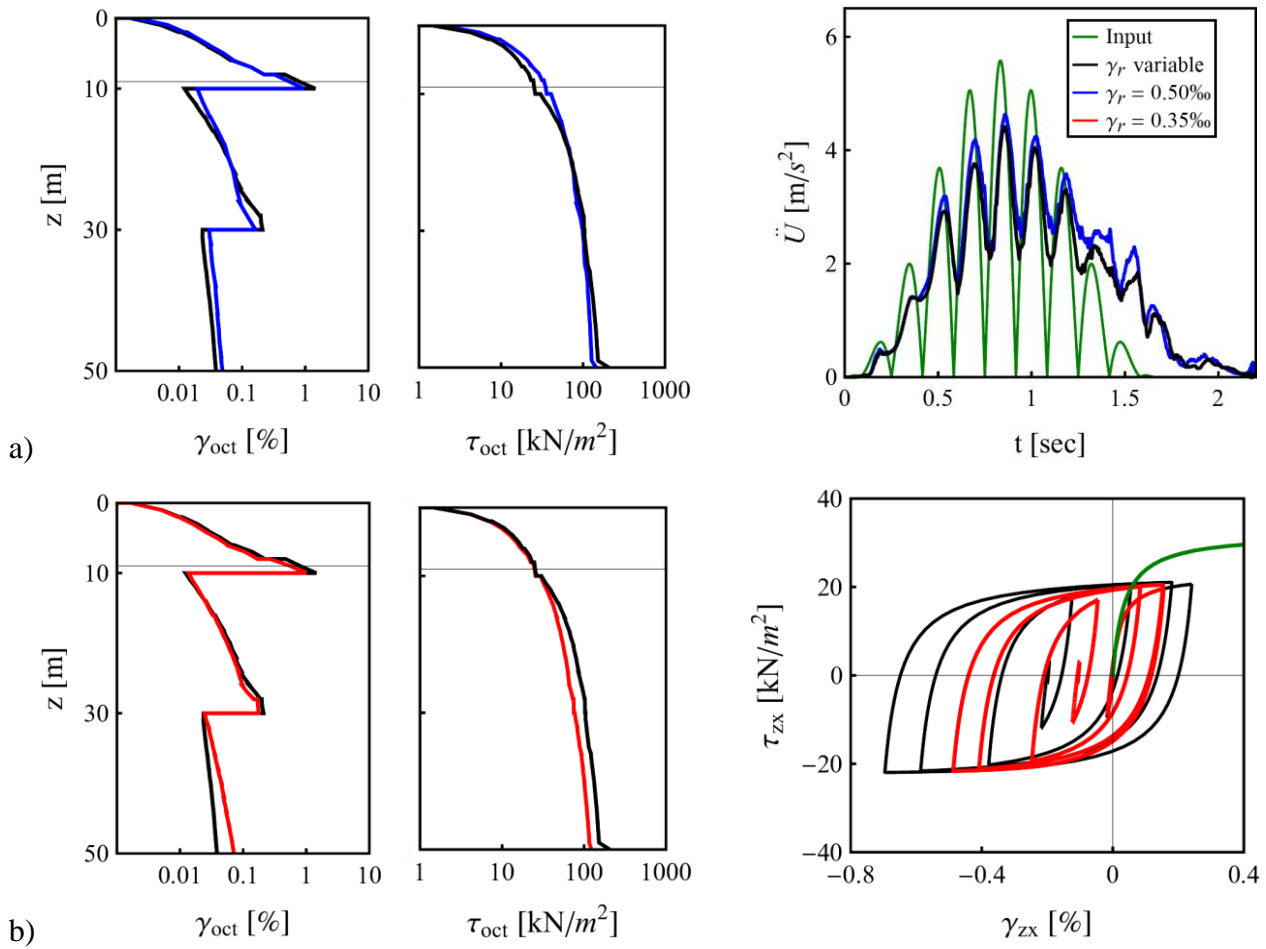

**Figure 12.** Comparison between the dynamic response to a three-component input signal of two soil profiles (B type) with homogeneous and variable reference shear strain. a) Maximum octahedral strain and stress profiles (left) and free surface acceleration time history compared with the bedrock acceleration (right); b) Maximum octahedral strain and stress profiles (left) and shear stress-strain loop at 9m depth (right).

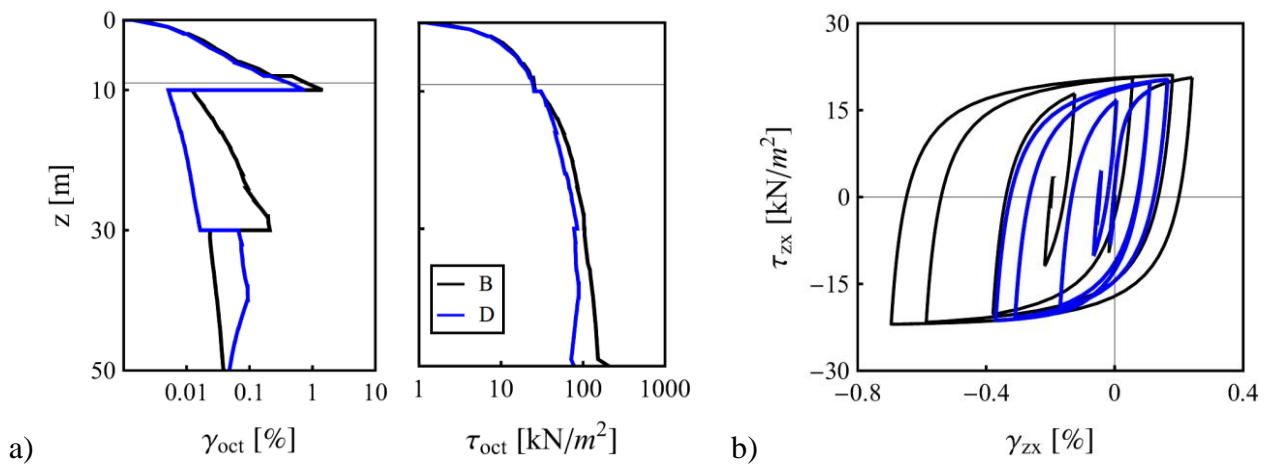

**Figure 13.** Influence of a stiff middle layer: dynamic response to a three-component input signal of soil profiles B and D with variable reference shear strain. a) Maximum octahedral strain and stress profiles; b) Shear stress-strain loop at 9m depth.